\documentclass[manuscript,screen]{acmart}

\usepackage[compact]{titlesec}         % you need this package
\titlespacing{\section}{0pt}{0pt}{0pt} % this reduces space between (sub)sections to 0pt, for example
\AtBeginDocument{%                     % this will reduce spaces between parts (above and below) of texts within a (sub)section to 0pt, for example - like between an 'eqnarray' and text
  \setlength\abovedisplayskip{0pt}
  \setlength\belowdisplayskip{0pt}}
  
\AtBeginDocument{%
  \providecommand\BibTeX{{%
    \normalfont B\kern-0.5em{\scshape i\kern-0.25em b}\kern-0.8em\TeX}}}

\setcopyright{acmcopyright}
\copyrightyear{2023}
\acmYear{2023}
\acmDOI{XXXXXXX.XXXXXXX}

\usepackage{graphicx}
\usepackage{commands}
\usepackage{mathtools}
\usepackage{hyperref}
\usepackage{algorithm2e}
\usepackage[colorinlistoftodos]{todonotes}
\usepackage{subcaption}
\usepackage{multirow}
\usepackage{enumitem}

\begin{document}

\title{AdBooster: Personalized Ad Creative Generation using Stable Diffusion Outpainting}
 
\author{Veronika Shilova}
\email{v.shilova@criteo.com}
\affiliation{%
 \institution{Criteo}
 \streetaddress{32 Blanche}
 \city{Paris}
 \country{France}}

\author{Ludovic Dos Santos}
\email{l.dossantos@criteo.com}
\affiliation{%
 \institution{Criteo}
 \streetaddress{32 Blanche}
 \city{Paris}
 \country{France}}

\author{Flavian Vasile}
\email{f.vasile@criteo.com}
\affiliation{%
 \institution{Criteo}
 \streetaddress{32 Blanche}
 \city{Paris}
 \country{France}}

\author{Gaëtan Racic}
\email{g.racic@criteo.com}
\affiliation{%
 \institution{Criteo}
 \streetaddress{32 Blanche}
 \city{Paris}
 \country{France}}

\author{Ugo Tanielian}
\email{ugo.tanielian@criteo.com}
\affiliation{%
 \institution{Criteo}
 \streetaddress{32 Blanche}
 \city{Paris}
 \country{France}}
 

\begin{abstract}
    In digital advertising, the selection of the optimal item (recommendation) and its best creative presentation (creative optimization) have traditionally been considered separate disciplines. However, both contribute significantly to user satisfaction, underpinning our assumption that it relies on both an item's relevance and its presentation, particularly in the case of visual creatives. In response, we introduce the task of {\itshape Generative Creative Optimization (GCO)}, which proposes the use of generative models for creative generation that incorporate user interests, and {\itshape AdBooster}, a model for personalized ad creatives based on the Stable Diffusion outpainting architecture. This model uniquely incorporates user interests both during fine-tuning and at generation time. To  further improve AdBooster's performance, we also introduce an automated data augmentation pipeline. Through our experiments on simulated data, we validate AdBooster's effectiveness in generating more relevant creatives than default product images, showing its potential of enhancing user engagement.
\end{abstract}




\maketitle

\section{Introduction}

Creative optimization seeks to identify the most suitable message for its audience, focusing on the aesthetics and the efficient presentation of the information rather than on the content itself. As such, it complements recommendation systems by addressing not the question of {\itshape what to display} to the user, but the question of {\itshape how to display} it.

Traditional recommendation literature assumes that the reward received by the recommender system from the user (e.g. click, conversion) is solely due to the quality of the recommendation. This marginalizes out the impact of the creative optimization step, that selects the best image together with its text description to describe the recommended item. We believe this is a missed opportunity since the two steps should be in sync. As an illustrative example, imagine that the user intent is {\itshape sport shoes for work}, and that the recommender system returns a very relevant pair of sneakers, but the creative optimization step selects the best non-personalized creative for the product, which is the {\itshape pair of shoes worn on a basketball terrain}. It is easy to imagine that the choice of such a creative will affect the final probability of conversion on the said pair of sneakers.

However, such a limitation seems almost insurmontable given the current creative optimization approaches:
currently, many of the performance advertising companies employ {\itshape Contextual Multiarmed Bandits} as a solution for creative optimization. This approach is particularly fitting in the cases where the range of alternatives for a specific item is limited, usually ranging between 3-10 ad versions and when the number of personalisation contexts is also small.

Very recently, the field of image and text generation witnessed a surge of advances (such as the open release of the Stable Diffusion model \cite{rombach2022high}), that lead to photorealistic {\itshape Text-to-Image Generation} capabilities. In our view, this presents the field of ad creative optimization with a tremendous opportunity to reinvent itself and get itself perfectly aligned with the recommendation step, leading to possibly significant performance improvements.

\subsection{Our proposal}

In the domain of digital advertising, one major challenge is to create ads that can resonate with diverse user interests. The creative element of an ad has a profound impact on the user's propensity to engage, making personalized creative content an essential factor in ad success. Existing approaches to ad creation, while effective to some extent, fall short in tailoring the creatives to individual user's interests. Our work attempts to bridge this gap by introducing and formalizing the task of {\itshape Generative Creative Optimization (\emph{GCO})}, a novel concept that leverages user interest signals to personalize creative generation.

We present {\itshape AdBooster}, a new model for generating personalized ad creatives based on the Stable Diffusion outpainting model architecture. AdBooster offers a distinctive way to incorporate individual user interest distribution during the fine-tuning process and at creative generation time, thereby ensuring that the creatives generated are specifically tailored to each user's preferences. To the best of our knowledge, this is the first study that integrates a generative model with ad creative personalization based on user interests.

In addition, we propose an innovative automated data augmentation pipeline to enhance the robustness and performance of our model. This pipeline feeds into the fine-tuning process of AdBooster, adding another layer of optimization to the creative generation process.

We carry out a series of experiments to demonstrate the efficacy of our approach. The results showcase an increased relevance of the ad creatives generated by AdBooster compared to the default product images, thereby leading to improved user engagement. The AdBooster model, with its novel Generative Creative Optimization approach, automated data augmentation pipeline, and Stable Diffusion outpainting architecture, holds significant potential in revolutionizing the ad creative generation process.

Contributions:
\begin{itemize}
    \item Introduce and formulate the task of {\itshape Generative Creative Optimization} that personalizes the creative generation using the same  user interest signal with the recommendation step, therefore perfectly aligning the two;
    \item Propose {\itshape AdBooster}, a generative ad creative model based on the Stable Diffusion outpainting model architecture that can be fine-tuned on the user interests distribution and be conditioned on the user interests at creative generation time;
    \item Propose an automated data augmentation pipeline that can be used for fine-tuning the AdBooster model;
    \item Run extensive experiments that show the increased relevance of generative ad creatives compared to default product images.
\end{itemize}
\section{Related work}
\subsection{Creative Optimization}

Creative optimization, a vital component in digital advertising, has remained relatively unexplored in machine learning literature, often overshadowed by the extensive research focused on recommendation systems \cite{bobadilla2013recommender, zhang2019deep}. Creative optimization targets the essential question of 'how' to present the recommended item, which significantly contributes to user engagement and satisfaction \cite{krajbich2010visual, teixeira2012determination}.

The \emph{Contextual Multiarmed Bandits} method has been widely employed for creative optimization, especially when the creative alternatives for a specific item are limited \cite{li2010contextual, lu2010contextual}. This technique adeptly balances exploration and exploitation, optimizing user engagement.

\subsection{Text-to-Image approaches}

Recently, with the advent of advanced image and text generation models, such as the Stable Diffusion model \cite{rombach2022high}, new avenues have been opened in creative optimization. These models, capable of generating photorealistic images from text descriptions, offer opportunities for personalized and dynamic creative generation.

The concept of diffusion probabilistic model (DM), a class of latent variable models inspired by  nonequilibrium thermodynamics, was introduced in \cite{sohl2015deep}. This architecture was later improved by Denoising Diffusion Probabilistic Model (DDPM) \cite{ho2020denoising} which offered a different training procedure based on a novel connection between DMs and denoising score matching with Langevin dynamics. Despite the high quality results on image generation, these models had extremely low inference speed. The issue was partially mitigated by using such advanced training and sampling strategies as Denoising Diffusion Implicit Model (DDIM) \cite{song2020denoising}, score-based diffusion \cite{song2020score}, and progressive distillation \cite{salimans2022progressive}. These architectures primarily use U-net \cite{ronneberger2015u} as their backbone neural network.

However, since all these methods operate directly in image pixel space, they still have a significant disadvantage of high training costs. To minimize the computational resources needed for training a diffusion model, inspired by the concept of latent image \cite{esser2021taming}, the Latent Diffusion Model (LDM) approach was introduced in \cite{rombach2022high}. Over time, this approach was expanded and refined into what is now known as Stable Diffusion.

Furthermore, it is worth noting that the mathematical concepts behind diffusion models, such as Langevin dynamics, score matching, SDEs, etc., make it hard to enter the subject. Therefore, some efforts have been taken in order to reformulate DMs in a simpler way. In \cite{heitz2023iterative}, the authors derive a minimalist yet powerful deterministic diffusion model that is simple to implement, numerically stable and achieves high quality results in a number of experiments.

Diffusion models have proven themselves effective in text-to-image generation tasks, producing state-of-the-art image generation results. This is typically accomplished by encoding text inputs into latent vectors using pre-trained language models such as CLIP \cite{radford2021learning}. There is a number of recent successful methods for text-to-image translation: Glide \cite{nichol2021glide} is a text-guided diffusion model that supports both image generation and editing; Stable Diffusion represents a large-scale implementation of latent diffusion aimed at achieving text-to-image generation; Imagen \cite{saharia2022photorealistic} is a text-to-image framework that forgoes latent images and instead directly diffuses pixels using a pyramid structure. Another approach called Dreambooth \cite{ruiz2023dreambooth} was built on top of Imagen to enable "personalization" of text-to-image diffusion models. Given as input a few images of a subject, it is able to generate new photorealistic images of it contextualized in different scenes.

Lately, two novel architectures were introduced in the field of large diffusion models: ControlNet \cite{zhang2023adding} and Composer \cite{huang2023composer}. They both tackle the problem of limited controllability of large diffusion models, allowing DMs to support additional input conditions, such  as edge maps, segmentation maps, keypoints, color histogram and more. By enhancing the control of large diffusion models, these advancements pave the way for broader applications and possibilities.

In this paper, we attempt to show how the recent breakthroughs in generative AI can be leveraged in the field of personalized advertisements generation, taking as an example fashion and retail domain specifically. This area has a lot of potential challenges and limitations. Ad creative algorithm would need to model complex fashion scenes and diverse background styles, while handling fine details, texture, complex object interactions and semantic consistency in the synthesized backgrounds.

%\begin{itemize}
    %\item \href{https://arxiv.org/pdf/2208.01618.pdf}{Textual Inversion}\cite{gal2022image}
   % \item \href{https://keras.io/examples/generative/random_walks_with_stable_diffusion/}{Latent Space Walking}
  % \item \href{https://arxiv.org/pdf/2204.06125.pdf}{DALLE2} \cite{ramesh2022hierarchical}
 % \item \href{https://lilianweng.github.io/posts/2021-07-11-diffusion-models/}{Lilian Weng's post on Diffusion models}
%\end{itemize}

\section{Our approach} 

In this section, we introduce the formal definitions of  recommendation and classical creative optimization and propose a Generative Creative Optimization objective, that uses generative text-to-image models to align recommendation and creative optimization by fully sharing the user context.

\subsection{Formal problem statement}

\paragraph{Recommendation}

The training objective for recommendation can be written as:

\begin{equation}
    \theta^* = \arg\max_{\theta} \sum_{(u,i)} R(u,i)\pi_\theta(i|u)
\end{equation}
where $R(u,i)$ is the reward that the user $u$ returns when shown a recommendation for item $i$ and $\pi_\theta$ is the recommendation policy that is represented for generality purposes as a distribution over all possible recommendable items. In most cases the policy will be a Dirac on the best item $i^*$.

At inference time, for each user u, we select the best item i using the learnt policy:
\begin{equation}
    i^* = \arg\max_{i} \pi_{\theta^*}(i | u)
\end{equation}

\paragraph{Classical Creative Optimization}
Creative testing is the way marketers compare the performance of different creatives in a campaign in order to evaluate which creatives yield better results. In practice marketers use two setups for matching creatives with user groups:  A/B testing, which is a commonly used method and widely known method, and the multi-armed bandit, which is less commonly known and used, usually due to its relative complexity. 

Formally, this can be seen as finding the policy that best matches product image creatives with the users. As a result, the training objective for creative optimization can be written as:

\begin{equation}
    \gamma^* = \arg\max_{\gamma} \sum_{(u,i)} R(u,\hat{c}(i))\pi_\gamma(\hat{c}(i)|u,i)
\end{equation}
where $\pi_\gamma$ is the creative optimization policy and $\hat{c}(i) \in C_i$ is one of the ad creatives available for product i.

At inference time, for each user u and the best item $i^*$, we select the best creative using the creative optimization policy:
\begin{equation}
    c^*(i^*) = \arg\max_{c(i^*)} \pi_{\gamma^*}(c(i^*) | u',i^*)
\end{equation}
Note: In practice, this is not conditioned on the full user context $u$, but on a low-granularity representation of it, $u'$. In our proposal, we will be able to condition the creative on the full user representation, allowing us to align the recommendation and creative optimization models.

\paragraph{Generative Creative Optimization}
With the advent of photorealistic image generation models, we are not  constrained anymore to the task of matching users with a discrete set of pre-existing creative choices and instead are able to generate an unbounded number of potential choices, making it a continuous optimization problem. As a result, the training objective for generative creative optimization task becomes:

\begin{equation}
    \gamma^* = \arg\max_{\gamma} \sum_{(u,i)} R(u,G_\gamma(c(i),u))
\end{equation}

where $G_\gamma$ is the creative generation network and $c(i)$ is the default product image (e.g. the product on a simple white background).

At inference time, for each user u and the best item $i^*$, we generate the best creative using the creative generation network that takes the same user representation $u$ as recommendation:
\begin{equation}
    c^*(i^*) = G_{\gamma^*}(c(i^*),u)
\end{equation}

If we explicitly separate the text and image descriptions of the initial asset, we have that: 

\begin{equation}
    \gamma^* = \arg\max_{\gamma} \sum_{(u,i)} R(u,G_\gamma(c(i),t(i),u))
\end{equation}

\subsection{Generative Creative Optimization using Stable Diffusion Outpainting}
In our paper we concentrate on the concrete problem of generating ad creatives using Stable Diffusion outpainting models. Given an outpainting model conditioned on image, text and product mask we have that:
\begin{equation}
    \gamma^* = \arg\max_{\gamma} \sum_{(u,i)} R(u,G_\gamma(c(i),T_\phi(i,u),m(i)))
\end{equation}

where:
\begin{itemize}
    \item $c(i)$ is the initial product image
    %\item $c'(i)$ is the generated product image \ludo{is it used somewhere?}
    \item $T_\phi(i,u)$ (resp. $T_\phi(i)$) is a text generative model that takes as inputs both the product and user interest descriptions (resp. the product descriptions) and returns the textual prompt for the image out-painting model
    \item $m(i)$ is the product image mask
\end{itemize}

\paragraph{Studying the impact of personalized generation using pretrained text and image generators}

For the rest of the paper we concentrate on studying the impact of contextualizing the generation of the images based on the user interests/type.
More explicitly, we are interested in showing that the quality of the generations gets better if contextualized with the full user context versus their un-contextualized counterparts:

\begin{equation}
   \sum_{(u,i)} R(u,G_\gamma(c(i),T_\phi(i,u),m(i)) > \sum_{(u,i)} R(u,G_\gamma(c(i),T_\phi(i),m(i))
\end{equation}

\subsection{Fine-tuning the outpainting model for fashion-specific ad creative generation}
\label{subsection:fine-tuning}
% \ludo{Don't we want to give more details since this is the core of the training procedure? Or do we want to focus on the GCO framework and the experiments only?}

We need to fine-tune Stable Diffusion (SD) \cite{rombach2022high} model to be able to use it efficiently for ad creative generation. It is essential, since SD was trained for inpainting task (masks of small size were used), whereas, in our case, we want to outpaint rather large regions. Furthermore, it is crucial in our task that the outpainting model does not modify the product itself. We noticed that without the fine-tuning, SD often extends the frontiers of a product.

A Stable Diffusion \cite{rombach2022high} model can be decomposed into several key models:
    \begin{itemize}
        \item A text encoder that projects the input prompt to a latent space.
        \item A variational autoencoder (VAE) that projects an input image to a latent space.
        \item A diffusion model that refines a latent vector and produces another latent vector, conditioned on the encoded text prompt.
        \item A decoder that generates images given a latent vector from the diffusion model.
    \end{itemize}

Only the diffusion model parameters are updated during fine-tuning, while the (pretrained) text and the image encoders are kept frozen. Therefore, the fine-tuning loss is calculated between the predicted noise by the diffusion model and the original noise.

To fine-tune SD for background outpainting and then to test our hypothesises on ad creatives generation, we need a large dataset of images with corresponding masks and captions. However, usually such datasets are not available. That is why we propose a data augmentation pipeline which automatically generates product masks and textual prompts from a given dataset of images which can be later used for fine-tuning. 

The data augmentation pipeline consists of the following steps: $(1)$ data cleaning (see appendix \ref{appendix:pipeline}); $(2)$ product mask extraction; $(3)$ prompt generation. To get masks, we use background removal model based on U$^2$-Net introduced in \cite{qin2020u2}. To generate prompts, we apply BLIP model to every image in the dataset which was introduced in \cite{li2022blip}. Figure \ref{fig:data_pipe} gives an example from data augmentation pipeline. The further details are provided in appendix \ref{appendix:pipeline}.

\begin{figure}[h]
  \centering
  \includegraphics[width=0.7\linewidth]{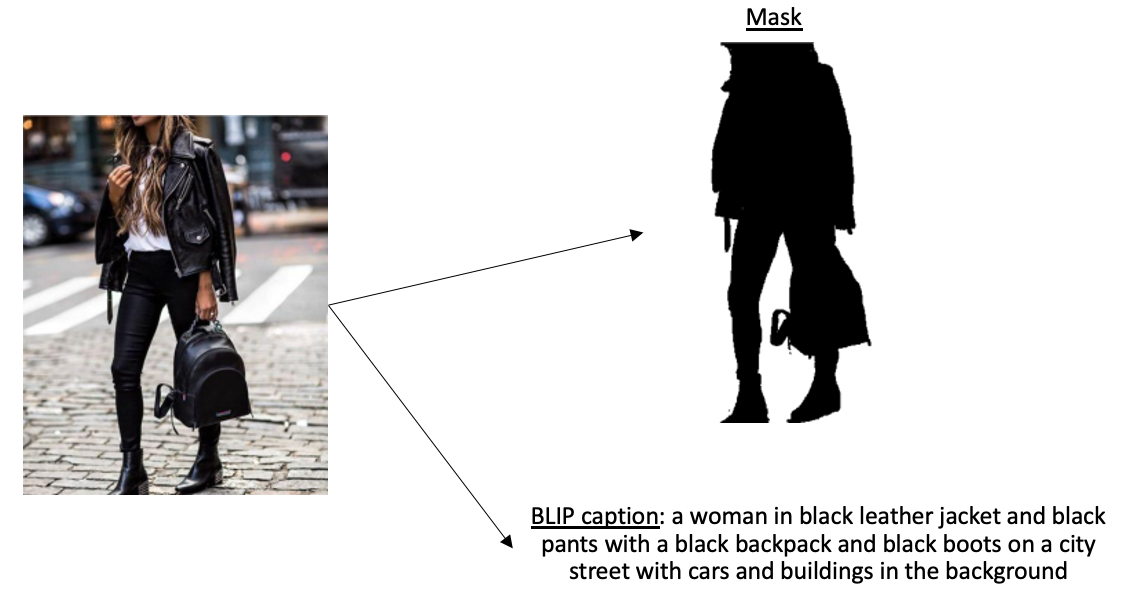}
  \caption{Example from data augmentation pipeline.}
     \label{fig:data_pipe}
\end{figure}

It is worth pointing out that it is also possible to add task specific regularization term to the loss function to prevent product extension. For example, after outpainting we can segment the product again and compare its mask with the one used for generation. However, in our experience, this step is not necessary and fine-tuning the model with data augmented pipeline is sufficient to overcome this potential drawback. 

The importance of fine-tuning is also confirmed by the decrease of Fréchet Inception Distance (FID) metric \cite{heusel2017gans}. On the dataset which are used in our experiments and which is introduced in section \ref{subsection:data}, we obtained FID of $30.46$ for pretrained SD model \cite{rombach2022high} and FID of $25.73$ for the fine-tuned model.

\subsection{User representation and prompting strategies}

Since one of the key ideas of this paper is to investigate the influence of ad personalization on users, we need to set a user representation first. A user $u$ is defined by a vector $z(u)$ which contains information about user's query $Q(u)$ (description of a product that a user is looking for) and a product context $C(u)$. For example, a user query can be "women red shoes" and its context may be "wedding". Here, we define $z_{QC}(u)$ as the concatenation of $Q(u)$ and $C(u)$ of a given user $u$, thus in the previous example, $z_{QC}(u)=$"women red shoes for wedding".

In our approach we generate the new image creative which is contextualized either directly on the user representation $z_{QC}(u)$, either on the textual prompt given by a text generative model $z_{P}(u)=T_{\phi}(i, u)$ which takes as input recommended product $i$, $Q(u)$ and $C(u)$. We use GPT-3.5 model (gpt-3.5-turbo) \cite{brown2020language} from OpenAI API as $T_{\phi}(i, u)$.

\section{Experiments}

In our experiments section, we highlight the impact of personalization on the perceived relevance of the items on the user feedback due to the shared user representation.

\subsection{Experimental setup}

In order to test our hypothesis we design the following experimental setup:
\begin{enumerate}
    \item We decide on an open product data set containing both product images and textual descriptions
    \item We generate synthetic user representations $z_{QC}(u)$ defined as the concatenation of shopping queries and product context
    %(e.g. "sports shoes for going out", "sport shoes for basketball")
    \item We generate a synthetic data set of (user, item) pairs by returning for every user a random product from the top k most similar products with $z_{QC}(u)$ in the text-based CLIP similarity space. This step aims at simulating a search/recommendation engine logic.
\end{enumerate}

\paragraph{The baselines}
The baseline creative for a product is the default catalog image (usually the product on a simple white background).
Additionally, since our test product images are actually not a white background we can calculate the uplift in performance versus a second baseline, which is the image of the product in its original background.

% \paragraph{Our approach}
% In our approach we generate the new image creative which is contextualized either directly on the user representation $z_{QC}(u)$, either on the textual prompt given by a text generative model $z_{P}(u)=T_{\phi}(i, u)$ which takes as input recommended product $i$, $Q(u)$ and $C(u)$. \ludo{how?}

\paragraph{Performance metric}
In order to compare our proposed approach and the baseline we propose to compute the reward obtained by the user seeing a product image creative as $R(u,i)=
<CLIP(z(u)),CLIP(c(i))>$, which can be seen as the probability of the user buying the item modeled as a function of the affinity between its stated interest and the product image.

In this experimental setup, we investigate two prompting strategies for outpainting: we generate an ad creative image using either $(1)$ $z_{QC}(u)$ or $(2)$ $z_{P}(u)$. In both cases, the performance metric is calculated with respect to initial user representation $z_{QC}(u)$, since we want to measure the influence of ad contextualization on a given user.

\subsection{Data}
\label{subsection:data}

For our experiments, we chose Pinterest's Shop the Look (STL) Dataset collected by \cite{kang2019complete}. The dataset contains around 60k scenes of people wearing fashion on different backgrounds ranging from various indoor settings (home, work, gym, shops etc.) to diverse outdoor ones (nature, city setups, etc.). We run STL dataset through the data augmentation described in section \ref{subsection:fine-tuning}.

Furthermore, for evaluation we generated 30 different user profiles with GPT-3.5 (gpt-3.5-turbo) model \cite{brown2020language} from OpenAI API, each containing user's query and context -- $z_{QC}$ and corresponding $z_{P}$. For the scope of this paper, we chose 4 different categories for backgrounds to experiment with. We mainly focus on seasons (winter, spring, summer, fall); settings (nature vs city setups); various life events (marriage, vacation, parties); and sports (tennis, basketball, hiking). 
% \ludo{the entire list in appendix and refers to it here}.

\subsection{Experimental results}

% \begin{table*}
%   \caption{Performance comparison (CLIP similarity score) between BestNeutralImage, BestContextualSetting, and Generative Creative Optimization (\emph{GCO}) across different contexts: Seasons, Settings, Events, and Sports}
%   \label{tab:perf_metric}
%   \begin{tabular}{ccccl}
%     \toprule
%     & BestNeutralImage & BestContextualSetting & \emph{GCO} ($z_{QC}$) $\boldsymbol{\uparrow}$ & \emph{GCO} ($z_{P}$) $\boldsymbol{\uparrow}$\\
%     \midrule
%     Seasons  & 22.903 & 21.925 & $\boldsymbol{25.103}$ & $\boldsymbol{24.801}$ \\
%     Settings & 20.11 & 18.879 & $\boldsymbol{25.617}$ & $\boldsymbol{25.094}$\\
%     Events   & 24.039 & 22.264 & $\boldsymbol{25.713}$  & $\boldsymbol{24.906}$\\
%     Sports   & 23.816 & 20.503 & $\boldsymbol{24.772}$ & $\boldsymbol{24.369}$\\ 
%   \bottomrule
% \end{tabular}
% \end{table*}

\begin{table*}
  \caption{Performance comparison (CLIP similarity score) between BestNeutralImage, BestContextualSetting, and Generative Creative Optimization (\emph{GCO}) across different contexts: Seasons, Settings, Events, and Sports}
  \label{tab:perf_metric}
  \begin{tabular}{ccccccc}
    \toprule
    & \multicolumn{2}{c}{} & \multicolumn{2}{c}{BestNeutralImage baseline} & \multicolumn{2}{c}{BestContextualSetting baseline}\\
    & BestNeutralImage & BestContextualSetting  & \emph{GCO} ($z_{QC}$) & \emph{GCO} ($z_{P}$) & \emph{GCO} ($z_{QC}$) & \emph{GCO} ($z_{P}$) \\
    \midrule
    % All contexts &   &  & $\boldsymbol{14.95\% \uparrow}$ & $\boldsymbol{12.53\% \uparrow}$\\
    Seasons  & 22.903 & 21.925  & $\boldsymbol{12.41\% \uparrow}$ & $\boldsymbol{10.54\% \uparrow}$& $\boldsymbol{15.56\% \uparrow}$ & $\boldsymbol{13.67\% \uparrow}$\\
    Settings & 20.11 & 18.879 & $\boldsymbol{14.68\% \uparrow}$ & $\boldsymbol{12.55\% \uparrow}$& $\boldsymbol{15.16\% \uparrow}$ & $\boldsymbol{12.96\% \uparrow}$\\
    Events   & 24.039 & 22.264 & $\boldsymbol{14.65\% \uparrow}$ & $\boldsymbol{10.45\% \uparrow}$& $\boldsymbol{15.34\% \uparrow}$  & $\boldsymbol{11.92\% \uparrow}$\\
    Sports   & 23.816 & 20.503 & $\boldsymbol{21.5\% \uparrow}$ & $\boldsymbol{20.37\% \uparrow}$& $\boldsymbol{27.87\% \uparrow}$ & $\boldsymbol{26.94\% \uparrow}$\\ 
  \bottomrule
\end{tabular}
\end{table*}

In Table \ref{tab:perf_metric} 
we compare two baselines: \emph{BestNeutralImage} which simulates the response rate we obtain by deploying a combination of content-based recommendation and no creative optimization, \emph{BestContextualSetting} which simulates the response rate we obtain by deploying a combination of content-based recommendation and full user-based creative optimization (which is an upper bound of the perf. of systems deployed in practice), and \emph{\emph{GCO}}, which blends content-based recommendation with a fully personalized creative, powered by our AdBooster model. In our setting, \emph{BestNeutralImage} is an image of a product on a white background (no user context at all), \emph{BestContextualSetting} is an image of the best product recommendation from the catalog (predefined context), and for \emph{\emph{GCO}}, it is an image generated via our AdBooster model (user context incorporated). 

% \ludo{merge information related to each "setting" in three sentences with what it simulates and the corresponding image. you can do a list.}

In our experiments, for each \emph{\emph{GCO}} $(z_{QC})$ and \emph{\emph{GCO}} $(z_{P})$, we generated 4 different images and kept the one with the maximum CLIP score for display and metrics calculation. Empirically, the images with the largest CLIP scores gave us the best perceptual quality in the most cases. However, it was not always true, and you can find some examples in the appendix \ref{appendix:additional_results}.

From the table \ref{tab:perf_metric}, we observe that scores for \emph{BestContextualSetting} are lower than for \emph{BestNeutralImage} on average. It can be explained by the fact that predefined ad images from a catalog do not always have the background that is aligned with user context. For example, potential user may look for "a yellow dress for the beach vacation", however, there may be no picture of a yellow dress in the catalog that is shot on background with a beach. This issue can be tackled with AdBooster model, since it can tailor background content to users' needs.

Quantitatively speaking, both \emph{\emph{GCO}} conditioned on $z_{QC}$ user representation and \emph{\emph{GCO}} conditioned on $z_{P}$ user representation give an uplift of CLIP score over the two baselines. \emph{\emph{GCO}} $(z_{QC})$ gives, on average, a $15\%$ increase in terms of CLIP similarity score across all categories. Moreover, \emph{\emph{GCO}} $(z_{P})$ also shows an increase in the CLIP similarity score, approximately by $13\%$.

\begin{figure}[h]
\centering
\begin{subfigure}[b]{\linewidth}
\includegraphics[width=\linewidth]{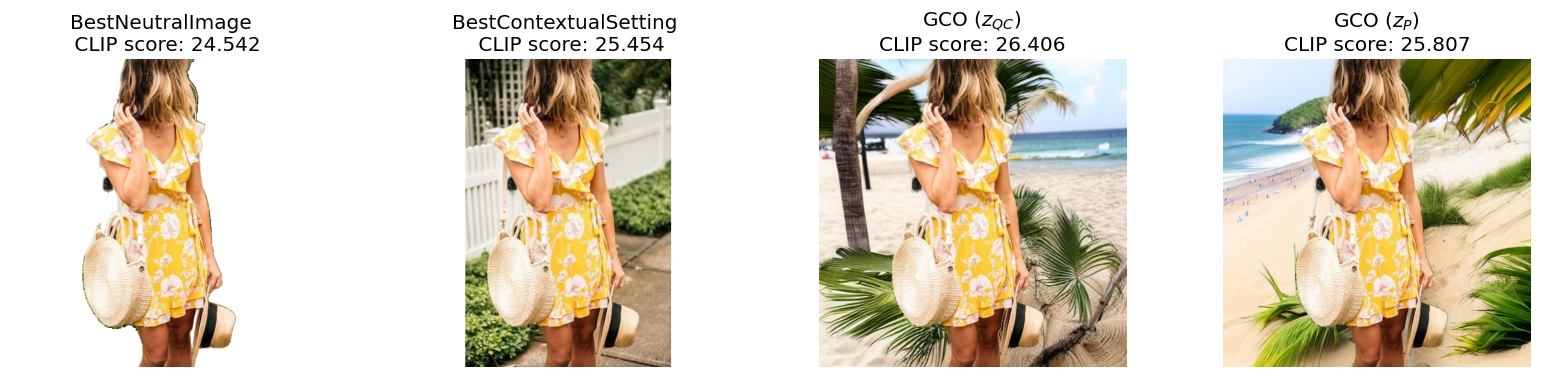}
\caption{Summer}\label{fig:summer_main}
\end{subfigure}

\begin{subfigure}[b]{\linewidth}
\includegraphics[width=\linewidth]{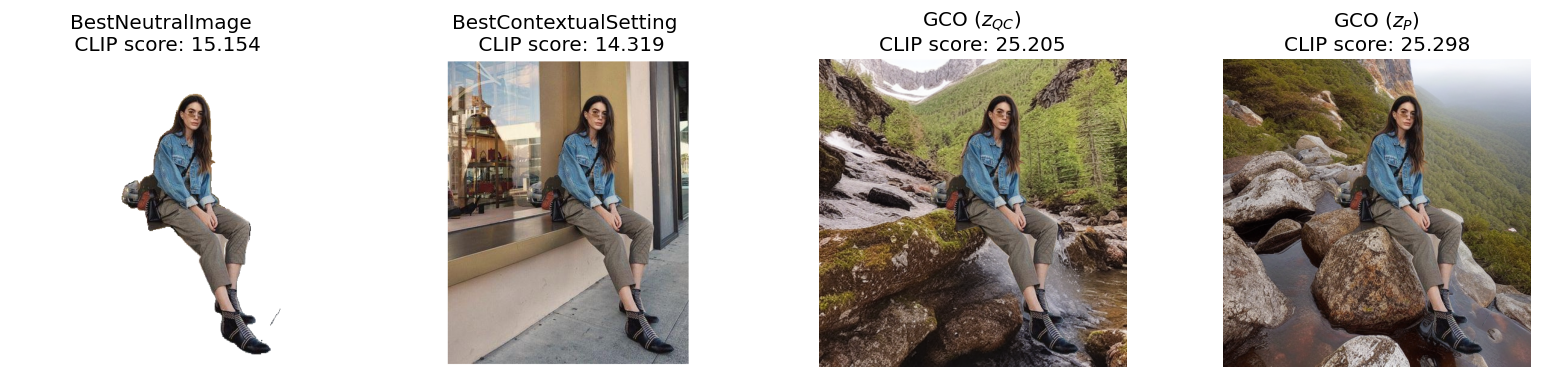}
\caption{Hiking}\label{fig:sport_main}
\end{subfigure}

\caption{Examples of AdBooster model generations compared to predefined product images in the seasons and sports categories. Corresponding $z_{QC}$ and $z_{P}$ are given in appendix \ref{appendix:additional_results}.}
\label{fig:settings_gen_ex_26674}
\end{figure}

In terms of qualitative analysis, we can conclude as well that generated backgrounds are closer to users' needs than the predefined ones, so the user is more likely to pay attention to the generated ads. Thus, the use of \emph{GCO} gives us a lot of flexibility regarding to ad creations.

In our experiments, we achieved the best visual results for seasons category. The outpainting model seems to capture almost perfectly "the idea" of natural settings and is able to reproduce them with a high degree of photo-realism. The city setups (buildings, streets) look very realistic as well. You can see some examples in figure \ref{fig:settings_gen_ex_26674}.

However, we still face a number of difficulties  that need to be resolved in the future. One of them is that we may occasionally get deduplicated generations (see appendix \ref{appendix:pipeline} for more information). Another minor model's struggle is that it cannot always generate reliably the light on the image or gets confused with ratio of object and subjects presented in the image. Moreover, generating people remains the most difficult task that even fine-tuning cannot resolve with sufficient realism. In the future work, we hope to resolve these problems by using and fine-tuning more powerful version of SD -- Stable Diffusion XL \cite{podell2023sdxl}.

Furthermore, we can compare the generation quality of \emph{GCO} ($z_{QC}$) against \emph{GCO} ($z_{P}$). Even though \emph{GCO} ($z_{QC}$) gives a slightly greater uplift in terms of CLIP score than \emph{GCO} ($z_{P}$), the latter show more variability in the generated background distribution. These generations are often more imaginative and detailed compared to the ones obtained with $z_{QC}$. All in all, simple concatenation is cheaper to use in practice, as you do not need to use an additional text generative model, and it already gives a sufficiently great uplift over two baselines. However, using generated prompts gives more flexibility for outpainting, for this reason, it might be interesting to further investigate different ways to do it.

% \begin{figure}[h]
%     \centering
%     \begin{subfigure}{.22\linewidth}
%         \centering
%         \includegraphics[width=\linewidth]{images/wb_26674.png}
%         \caption{BestNeutralImage}\label{fig:image1}
%     \end{subfigure}
%         \hfill
%     \begin{subfigure}{.22\linewidth}
%         \centering
%         \includegraphics[width=\linewidth]{images/original_26674.png}
%         \caption{BestContextualSetting}\label{fig:image2}
%     \end{subfigure}
%        \hfill
%     \begin{subfigure}{.22\linewidth}
%         \centering
%         \includegraphics[width=\linewidth]{images/z_concat_26674.png}
%         \caption{\emph{GCO} ($z_{concat}$)}\label{fig:image3}
%     \end{subfigure}
%      \hfill
%     \begin{subfigure}{.22\linewidth}
%         \centering
%         \includegraphics[width=\linewidth]{images/z_propmt_26674.png}
%         \caption{\emph{GCO} ($z_{prompt}$)}\label{fig:image4}
%     \end{subfigure}
    
%     \caption{Examples of AdBooster model generations compared to predefined product images in the seasons category.}
%     \label{fig:settings_gen_ex_26674}
% \end{figure}

You can find more examples of successful generations as well as some fails in appendix \ref{appendix:additional_results}.
\section{Conclusions}

The present study has made significant strides in the realm of personalized ad creative generation. By introducing the concept of Generative Creative Optimization (\emph{GCO}), we have opened new avenues for ad personalization that specifically target the creative aspect of ads based on user interest signals.

Our proposed model, AdBooster, utilizing the Stable Diffusion outpainting architecture, successfully demonstrates the effective integration of \emph{GCO}, exhibiting its capacity to generate personalized ad creatives. The unique fine-tuning process of AdBooster, which incorporates user interest distribution, is a groundbreaking approach to ad creative personalization.

The innovation of an automated data augmentation pipeline furthers the robustness and efficacy of our model, providing an additional dimension of optimization.

Furthermore, by using a shared user representation for both recommendation and creative optimization, we truly intertwine these tasks, aligning the goals of choosing the optimal item and crafting the ideal creative presentation. This joint approach creates synergy, enhancing the overall performance and effectiveness of digital advertising.

Through extensive experimental results, we have established the increased relevance of ad creatives generated by AdBooster compared to default product images, subsequently leading to improved user engagement.

In conclusion, our work presents a promising step towards revolutionizing the field of digital advertising. AdBooster, with its focus on personalization through \emph{GCO}, automated data augmentation, and Stable Diffusion outpainting, offers a compelling new method for ad creative generation that is both user-focused and efficient. We anticipate our findings to inspire further research and developments in this direction. As for the future steps, we plan to test Stable Diffusion XL \cite{podell2023sdxl} in our experiments to potentially further improve the visual quality of generated ads. We would also like to gather real user feedback for generated ad creatives to get more insight about the visual quality which might possibly be used to enhance \emph{GCO} performance. We also want to try to adopt such architectures as ControlNet \cite{zhang2023adding} and Composer \cite{huang2023composer} for our \emph{GCO} task to add more control over the ad creatives generation. Moreover, it is possible to investigate different prompting strategies to further boost generation flexibility.

\bibliographystyle{ACM-Reference-Format}
\bibliography{acmart}

\appendix
\section{Data augmentation pipeline}
\label{appendix:pipeline}

To further enhance our data augmentation pipeline, we also performed data cleaning step after masks' extraction. The idea was that in order to fully benefit from fine-tuning of SD, the following kind of images should be removed from the dataset 
\begin{itemize}
    \item images with simple, monochromatic backgrounds;
    \item images with deduplicated product (several different photos in one -- photo collage).
\end{itemize}

In order to remove images with monochromatic background, we thresholded on pixels intensities in the following way: 
\begin{itemize}
    \item convert image to grayscale;
    \item calculate standard deviation of background pixel intensities; 
    \item if the standard deviation is below a certain threshold, consider this image to have a monochromatic background. 
\end{itemize}

The standard deviation is a measure of how much the pixel intensities vary from the mean. If the standard deviation is low, it means that the pixel intensities are clustered around a single value, indicating a monochromatic background. The threshold is chosen empirically, in our experiments it equals to 20.

To remove deduplicated product images, we used the mask area in the following way:
\begin{itemize}
    \item calculate the ratio of masked area to the whole image area;
    \item if this ratio is above certain threshold, remove the image from the dataset.
\end{itemize}
This approach can also help to filter out too large, non-meaningful, masks which may be occasionally returned by the background removal algorithm. The threshold is chosen empirically, in our experiments it equals to 0.6.

\section{Additional results}
\label{appendix:additional_results}
\subsection{Successful generations}

\begin{figure}[h]
\centering
\begin{subfigure}[b]{\linewidth}
\includegraphics[width=\linewidth]{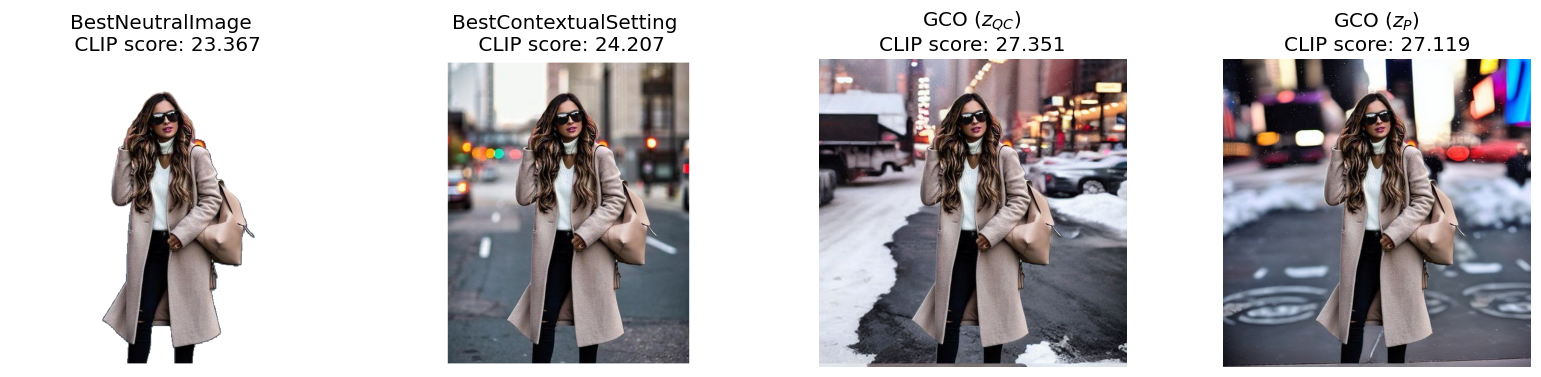}
\caption{Winter}\label{fig:winter}
\end{subfigure}

\begin{subfigure}[b]{\linewidth}
\includegraphics[width=\linewidth]{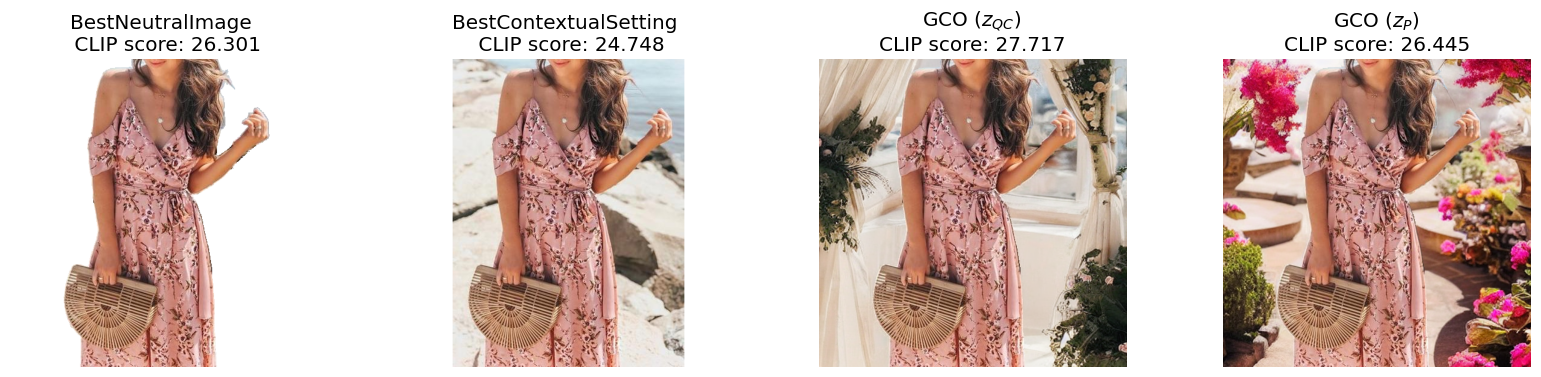}
\caption{Spring}\label{fig:spring}
\end{subfigure}

\begin{subfigure}[b]{\linewidth}
\includegraphics[width=\linewidth]{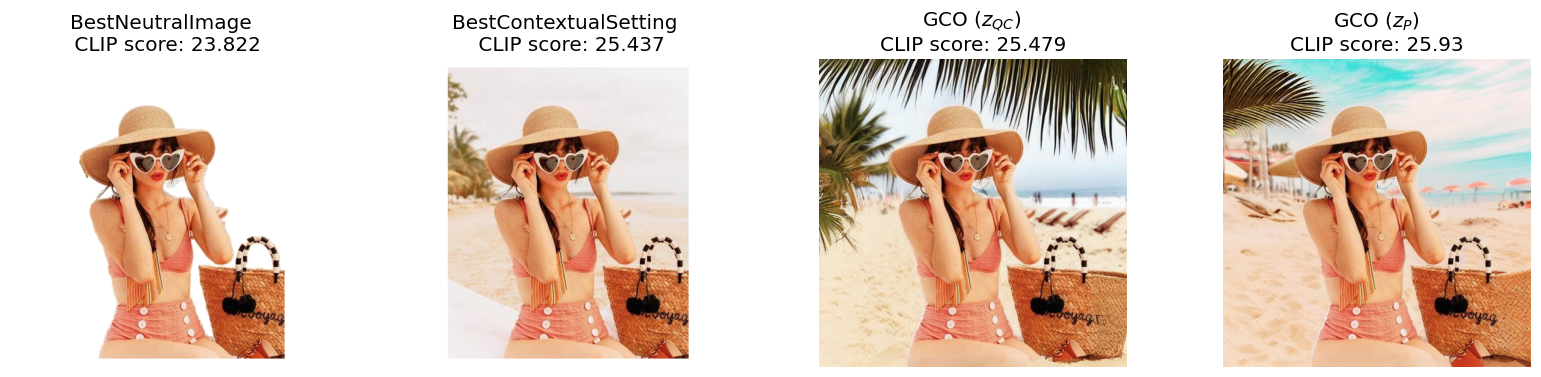}
\caption{Summer}\label{fig:summer}
\end{subfigure}

\begin{subfigure}[b]{\linewidth}
\includegraphics[width=\linewidth]{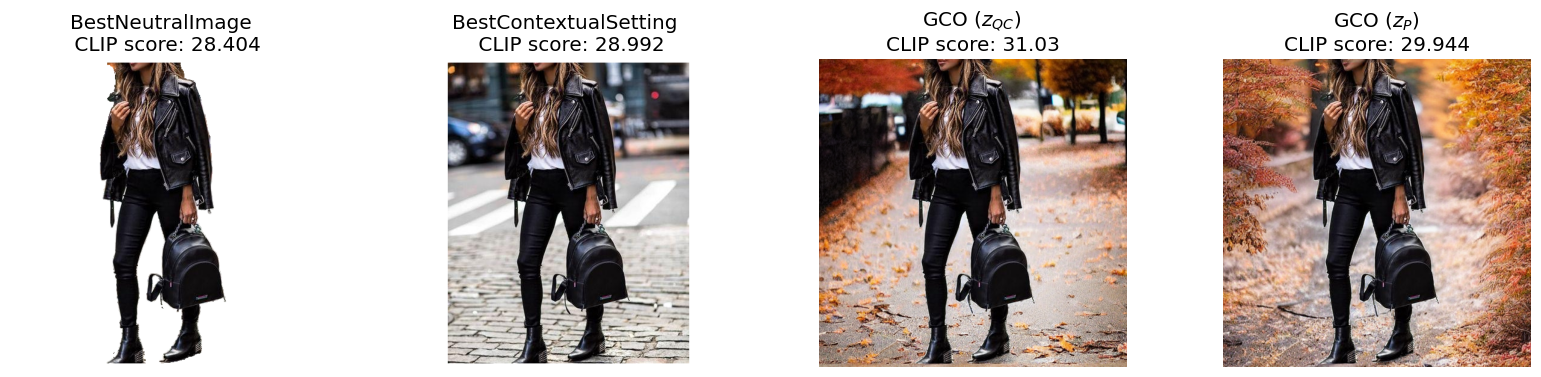}
\caption{Fall}\label{fig:fall}
\end{subfigure}
\caption{Examples of AdBooster model generations compared to predefined product images in the seasons category.}
\label{fig:seasons}
\end{figure}

\begin{figure}[h]
\centering
\begin{subfigure}[b]{\linewidth}
\includegraphics[width=\linewidth]{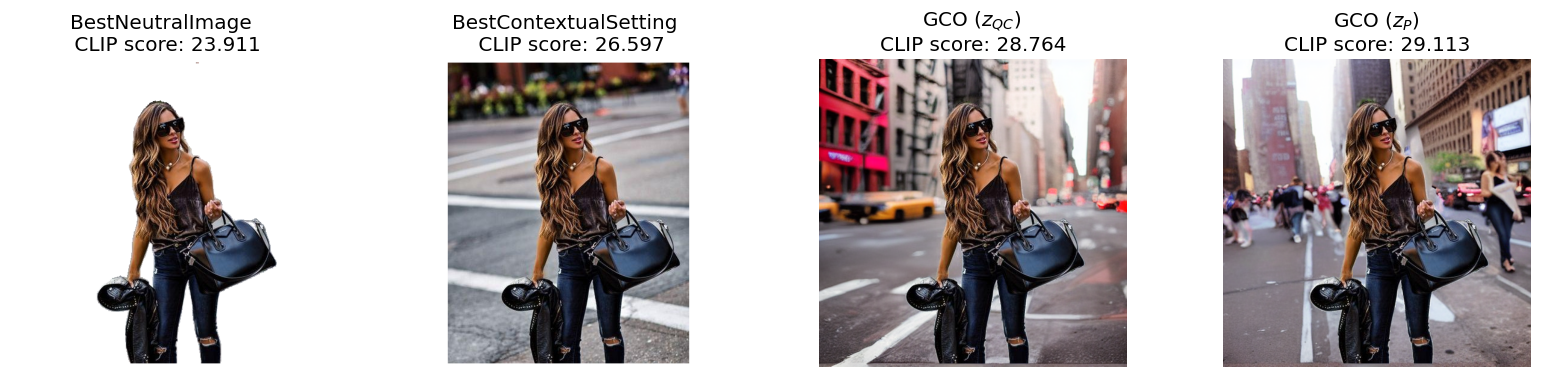}
\caption{City setup}\label{fig:city_woman}
\end{subfigure}

\begin{subfigure}[b]{\linewidth}
\includegraphics[width=\linewidth]{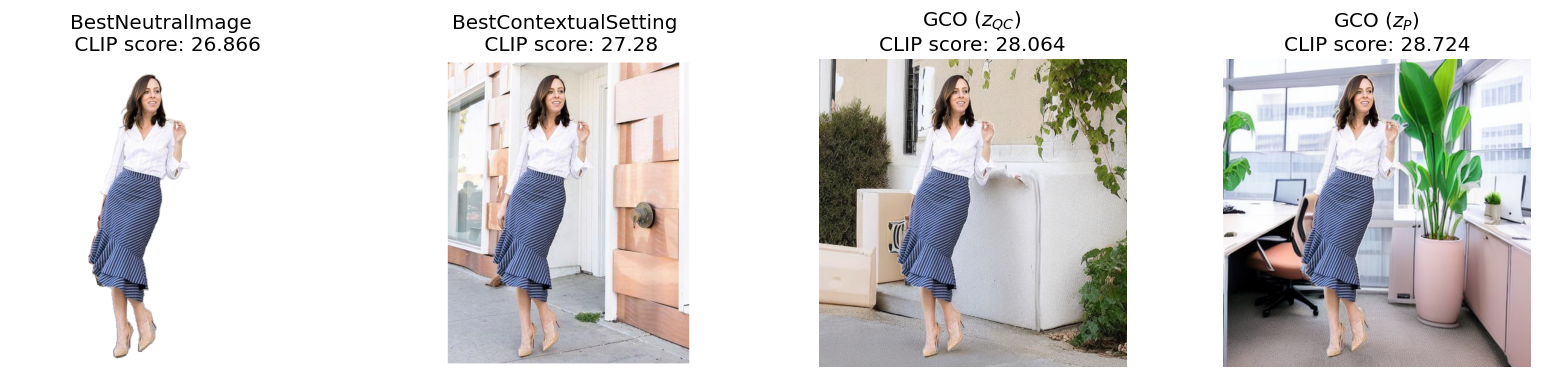}
\caption{City (office) setup}\label{fig:office_woman}
\end{subfigure}

\begin{subfigure}[b]{\linewidth}
\includegraphics[width=\linewidth]{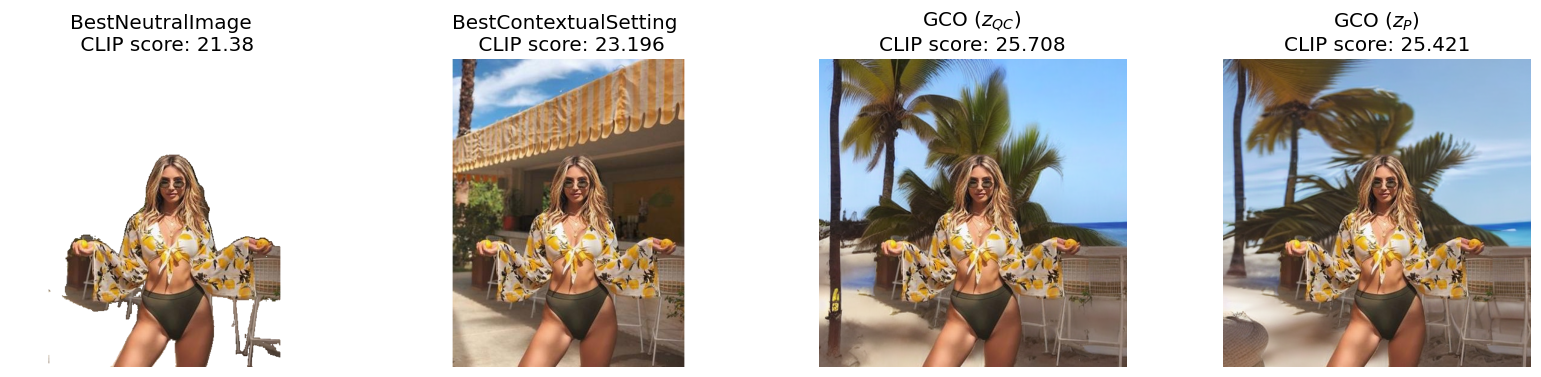}
\caption{Nature setup}\label{fig:nature1}
\end{subfigure}

\begin{subfigure}[b]{\linewidth}
\includegraphics[width=\linewidth]{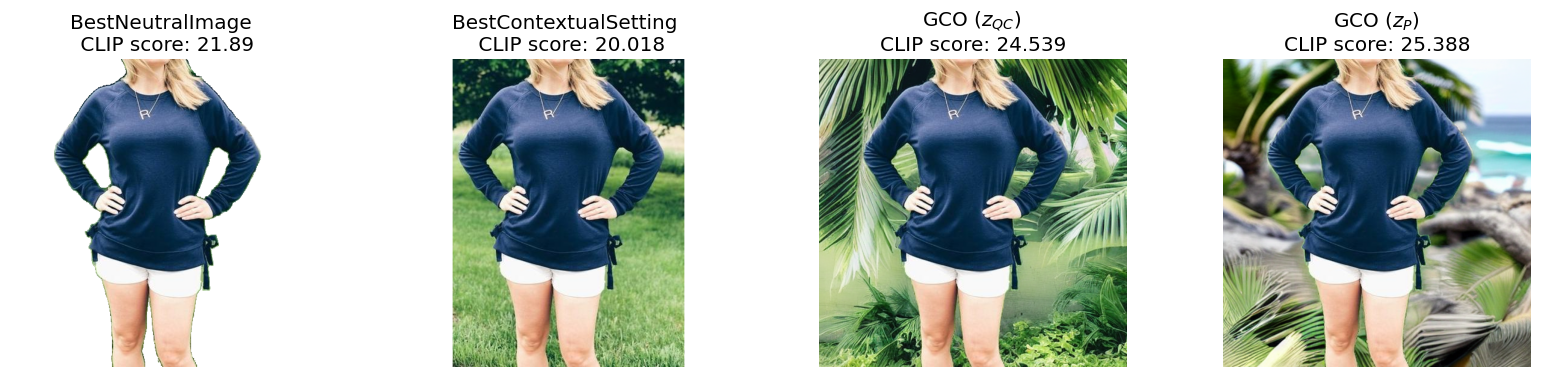}
\caption{Nature setup}\label{fig:nature2}
\end{subfigure}
\caption{Examples of AdBooster model generations compared to predefined product images in the settings category.}
\label{fig:settings}
\end{figure}

\begin{figure}[h]
\centering
\begin{subfigure}[b]{\linewidth}
\includegraphics[width=\linewidth]{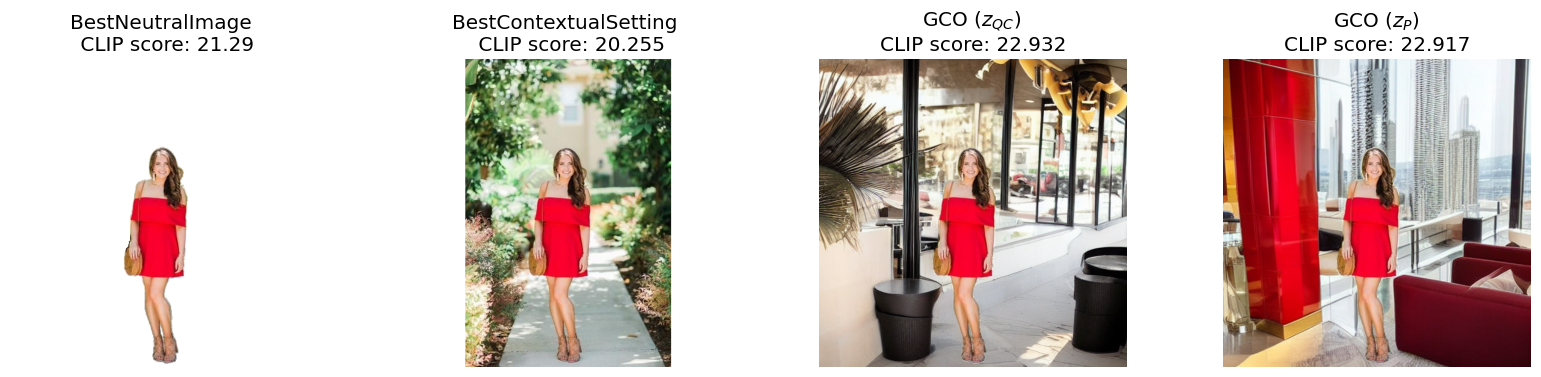}
\caption{Party}\label{fig:party}
\end{subfigure}

\begin{subfigure}[b]{\linewidth}
\includegraphics[width=\linewidth]{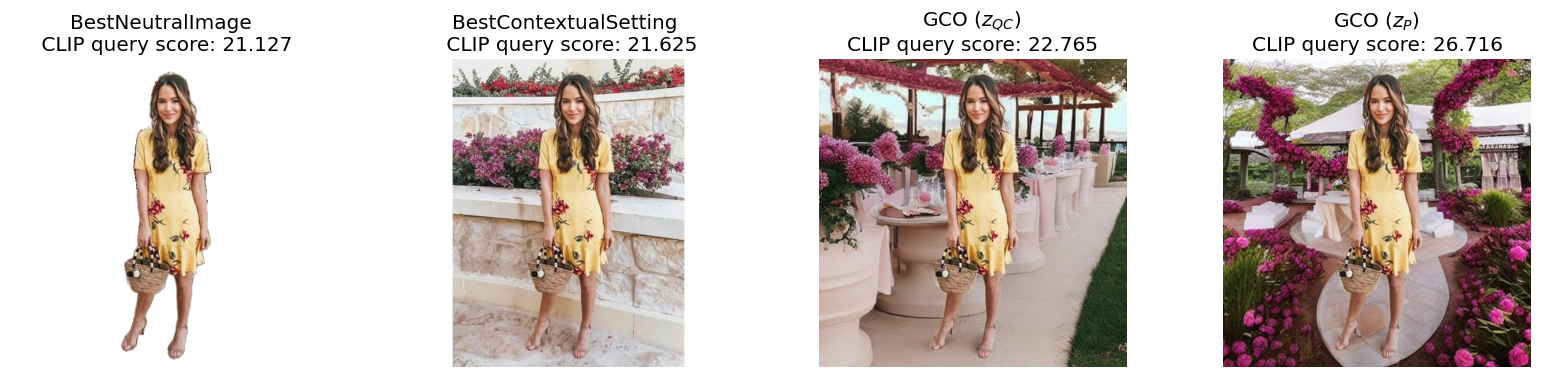}
\caption{Wedding}\label{fig:wedding}
\end{subfigure}

\caption{Examples of AdBooster model generations compared to predefined product images in the events category.}
\label{fig:events}
\end{figure}

\begin{figure}[h]
\centering
\begin{subfigure}[b]{\linewidth}
\includegraphics[width=\linewidth]{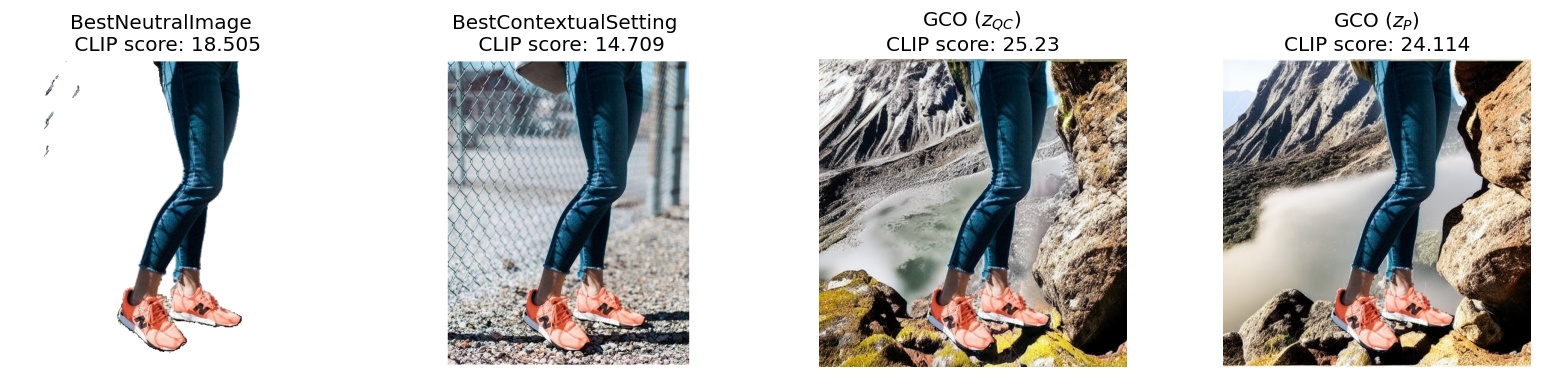}
\caption{Hiking}\label{fig:hiking}
\end{subfigure}

\begin{subfigure}[b]{\linewidth}
\includegraphics[width=\linewidth]{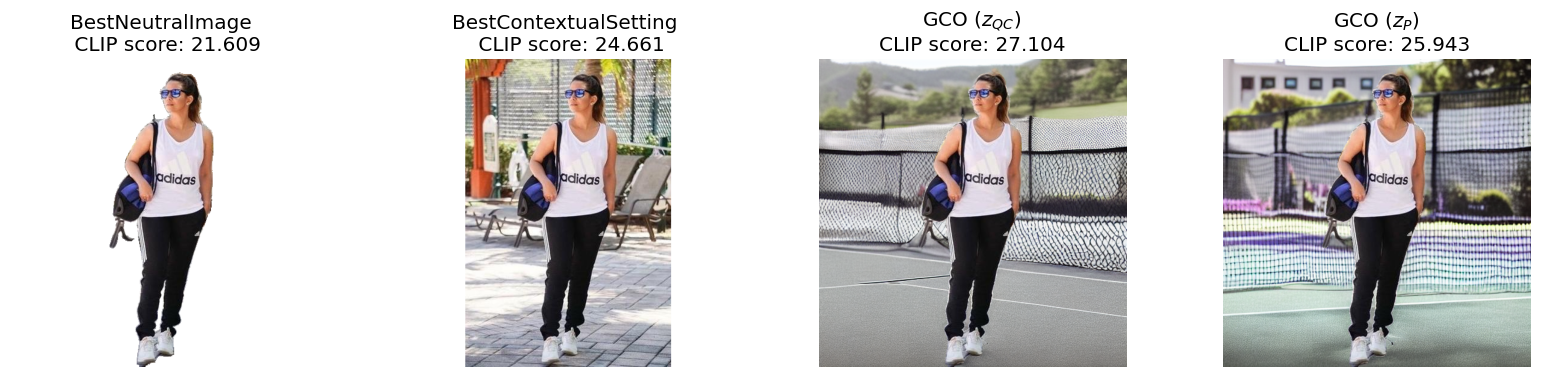}
\caption{Tennis}\label{fig:tennis}
\end{subfigure}

\caption{Examples of AdBooster model generations compared to predefined product images in the sports category.}
\label{fig:sports}
\end{figure}

Figures \ref{fig:settings_gen_ex_26674}, \ref{fig:seasons}, \ref{fig:settings}, \ref{fig:events} and \ref{fig:sports} give more examples of successful GCO generations. You can find the corresponding $z_{QC}$ (Query, Context) and $z_{P}$ (Prompt) below.

For figure \ref{fig:summer_main}:
\begin{itemize}
    \item \textbf{Query}: yellow summer dress for beach vacation;
    \item \textbf{Context}: planning a trip to Cancun in two weeks and needs a new dress for the beach;
    \item \textbf{Prompt}: A woman with her hair blowing in the wind stands on a sandy beach in a gorgeous yellow summer dress, with turquoise waves crashing behind her and the bright sun shining down, conveying the perfect beach vacation vibe.
\end{itemize}

For figures \ref{fig:sport_main}, \ref{fig:hiking}:
\begin{itemize}
    \item \textbf{Query}: waterproof hiking boots;
    \item \textbf{Context}: for a weekend hiking trip in the mountains;
    \item \textbf{Prompt}: A rugged pair of waterproof hiking boots sit atop a rocky outcropping overlooking the misty valley below, evoking a sense of adventure and the promise of breathtaking vistas on a weekend trek through the mountains.
\end{itemize}

For figure \ref{fig:winter}:
\begin{itemize}
    \item \textbf{Query}: women's long winter coat for snowy weather;
    \item \textbf{Context}: winter business trip to New York City;
    \item \textbf{Prompt}: A stylish woman wearing a long, black winter coat stands in Times Square as snow falls around her, holding a briefcase and looking confidently towards the camera.
\end{itemize}

For figures \ref{fig:spring}, \ref{fig:wedding}:
\begin{itemize}
    \item \textbf{Query}: elegant formal dress;
    \item \textbf{Context}: spring outdoor wedding guest;
    \item \textbf{Prompt}: A woman wearing a beautiful spring formal dress stands in a garden with blooming flowers, as she smiles at the camera as a guest of an outdoor wedding.
\end{itemize}

For figure \ref{fig:summer}:
\begin{itemize}
    \item \textbf{Query}: trendy sunglasses;
    \item \textbf{Context}: needed for a beach vacation this summer;
    \item \textbf{Prompt}: A vibrant photo featuring a stylish pair of sunglasses against the backdrop of a picturesque sandy beach with crystal clear turquoise waters and palm trees swaying in the gentle sea breeze.
\end{itemize}

For figure \ref{fig:fall}:
\begin{itemize}
    \item \textbf{Query}: black leather jacket;
    \item \textbf{Context}: for everyday use in the fall season;
    \item \textbf{Prompt}: A chic black leather jacket paired with distressed denim jeans stands out against the breathtaking backdrop of a colorful autumn forest.
\end{itemize}

% \begin{figure}[h]
%   \centering
%   \includegraphics[width=\linewidth]{images/user_0_prod_3_idx_45066.png}
%   \caption{Examples of AdBooster model generations compared to predefined product images in the seasons category.\\\hspace{\textwidth}
%   The corresponding user profile:\\\hspace{\textwidth}
%   Query: women's long winter coat for snowy weather\\\hspace{\textwidth}
% Context: winter business trip to New York City\\\hspace{\textwidth}
% Prompt: A stylish woman wearing a long, black winter coat stands in Times Square as snow falls around her, holding a briefcase and looking confidently towards the camera.}
%      \label{fig:seasons_gen_ex_45066}
% \end{figure}

% \begin{figure}[h]
%   \centering
%   \includegraphics[width=\linewidth]{images/user_4_prod_0_idx_30691.png}
%   \caption{Examples of AdBooster model generations compared to predefined product images in the seasons category.\\\hspace{\textwidth}
%   The corresponding user profile:\\\hspace{\textwidth}Query: black leather jacket\\\hspace{\textwidth}
% Context: for everyday use in the fall season\\\hspace{\textwidth}
% Prompt: A chic black leather jacket paired with distressed denim jeans stands out against the breathtaking backdrop of a colorful autumn forest.}
%      \label{fig:seasons_gen_ex_30691}
% \end{figure}

For figure \ref{fig:city_woman}:
\begin{itemize}
    \item \textbf{Query}: women’s jeans;
    \item \textbf{Context}: fashionable jeans for summer photoshoot in New York city streets;
    \item \textbf{Prompt}: A woman wearing stylish summer jeans posing confidently on a busy New York city street with towering skyscrapers and bustling crowds forming a dynamic and stunning background.
\end{itemize}

For figure \ref{fig:office_woman}:
\begin{itemize}
    \item \textbf{Query}: comfortable knee-length skirts for summer;
    \item \textbf{Context}: for an office with a business-casual dress code and need breathable skirts for the hot summer weather;
    \item \textbf{Prompt}: Woman in an office setting wearing stylish and comfortable knee-length skirts in various pastel colors, against the backdrop of a bright and airy conference room with large windows and potted plants.
\end{itemize}

For figure \ref{fig:nature1}:
\begin{itemize}
    \item \textbf{Query}: yellow sunglasses for beach vacation;
    \item \textbf{Context}: upcoming beach vacation, preferably polarized and scratch-resistant to protect eyes from the sun and the saltwater;
    \item \textbf{Prompt}: A vibrant and lively photo featuring a pair of yellow, polarized, scratch-resistant sunglasses well-suited for a beach vacation, set against a breathtaking background of clear blue skies, glistening waves, and soft sandy beaches stretching into the horizon.
\end{itemize}

For figure \ref{fig:nature2}:
\begin{itemize}
    \item \textbf{Query}: summer shorts for curvy women;
    \item \textbf{Context}: planning a beach vacation and wants comfortable shorts that flatter her curves;
    \item \textbf{Prompt}: A vibrant beach background with crystal blue ocean waves crashing on shore.
\end{itemize}

% \begin{figure}[h]
%   \centering
%   \includegraphics[width=\linewidth]{images/user_12_prod_4_idx_31178.png}
%   \caption{Examples of AdBooster model generations compared to predefined product images in the settings category.\\\hspace{\textwidth}
%   The corresponding user profile:\\\hspace{\textwidth}Query: women’s jeans\\\hspace{\textwidth}
% Context: fashionable jeans for summer photoshoot in New York city streets\\\hspace{\textwidth}
% Prompt: A women wearing stylish summer jeans posing confidently on a busy New York city street with towering skyscrapers and bustling crowds forming a dynamic and stunning background.}
%      \label{fig:settings_gen_ex_31178}
% \end{figure}

% \begin{figure}[h]
%   \centering
%   \includegraphics[width=\linewidth]{images/user_29_prod_0_idx_41441.png}
%   \caption{Examples of AdBooster model generations compared to predefined product images in the settings category.\\\hspace{\textwidth}
%   The corresponding user profile:\\\hspace{\textwidth}Query: men's casual summer shirts\\\hspace{\textwidth}
% Context: wear during hot summer days in the city\\\hspace{\textwidth}
% Prompt: The photo captures a group of stylish men wearing colorful and comfortable summer shirts while strolling through a bustling city street lined with tall buildings and lush green trees, with the warm sunshine shining down on them.}
%      \label{fig:settings_gen_ex_41441}
% \end{figure}

For figure \ref{fig:party}:
\begin{itemize}
    \item \textbf{Query}: red dress for a cocktail party;
    \item \textbf{Context}: attending a party at a fancy;
    \item \textbf{Prompt}: The photo depicts a stunning woman wearing a bright red dress, standing confidently in front of a luxurious modern lounge with a dazzling city skyline as the backdrop, exuding an air of elegance and sophistication.
\end{itemize}

% For figure \ref{fig:wedding}:
% \begin{itemize}
%     \item \textbf{Query}: elegant formal dress;
%     \item \textbf{Context}: spring outdoor wedding guest;
%     \item \textbf{Prompt}: A woman wearing a beautiful spring formal dress stands in a garden with blooming flowers, as she smiles at the camera as a guest of an outdoor wedding. 
% \end{itemize}

% \begin{figure}[h]
%   \centering
%   \includegraphics[width=\linewidth]{images/user_8_prod_1_idx_22219.png}
%   \caption{Examples of AdBooster model generations compared to predefined product images in the events category.\\\hspace{\textwidth}
%   The corresponding user profile:\\\hspace{\textwidth}Query: red dress for a cocktail party\\\hspace{\textwidth}
% Context: attending a party at a fancy lounge\\\hspace{\textwidth}
% Prompt: The photo depicts a stunning woman wearing a bright red dress, standing confidently in front of a luxurious modern lounge with a dazzling city skyline as the backdrop, exuding an air of elegance and sophistication.}
%      \label{fig:events_gen_ex_8613}
% \end{figure}

% For figure \ref{fig:hiking}:
% \begin{itemize}
%     \item \textbf{Query}: waterproof hiking boots;
%     \item \textbf{Context}: for a weekend hiking trip in the mountains;
%     \item \textbf{Prompt}: A rugged pair of waterproof hiking boots sit atop a rocky outcropping overlooking the misty valley below, evoking a sense of adventure and the promise of breathtaking vistas on a weekend trek through the mountains.
% \end{itemize}

For figure \ref{fig:tennis}:
\begin{itemize}
    \item \textbf{Query}: sunglasses for outdoor sports;
    \item \textbf{Context}: wear while playing tennis in the summer;
    \item \textbf{Prompt}: A vibrant and energetic photo captured a player enjoying a game of tennis in the bright summer sun, wearing trendy sunglasses perfectly complementing the colorful backdrop.
\end{itemize}

% \begin{figure}[h]
%   \centering
%   \includegraphics[width=\linewidth]{images/user_2_prod_0_idx_19531.png}
%   \caption{Examples of AdBooster model generations compared to predefined product images in the sports category.\\\hspace{\textwidth}
%   The corresponding user profile:\\\hspace{\textwidth}Query: waterproof hiking boots\\\hspace{\textwidth}
% Context: for a weekend hiking trip in the mountains\\\hspace{\textwidth}
% Prompt: A rugged pair of waterproof hiking boots sit atop a rocky outcropping overlooking the misty valley below, evoking a sense of adventure and the promise of breathtaking vistas on a weekend trek through the mountains.}
%      \label{fig:sports_gen_ex_19531}
% \end{figure}

% \begin{figure}[h]
%   \centering
%   \includegraphics[width=\linewidth]{images/user_25_prod_0_idx_8613.png}
%   \caption{Examples of AdBooster model generations compared to predefined product images in the sports category.\\\hspace{\textwidth}
%   The corresponding user profile:\\\hspace{\textwidth}Query: sunglasses for outdoor sports\\\hspace{\textwidth}
% Context:  wear while playing tennis in the summer\\\hspace{\textwidth}
% Prompt: A vibrant and energetic photo captured a player enjoying a game of tennis in the bright summer sun, wearing trendy sunglasses perfectly complementing the colorful backdrop.}
%      \label{fig:sports_gen_ex_8613}
% \end{figure}

\subsection{Failed generations}

\begin{figure}[h]
\centering
\begin{subfigure}[b]{\linewidth}
\includegraphics[width=\linewidth]{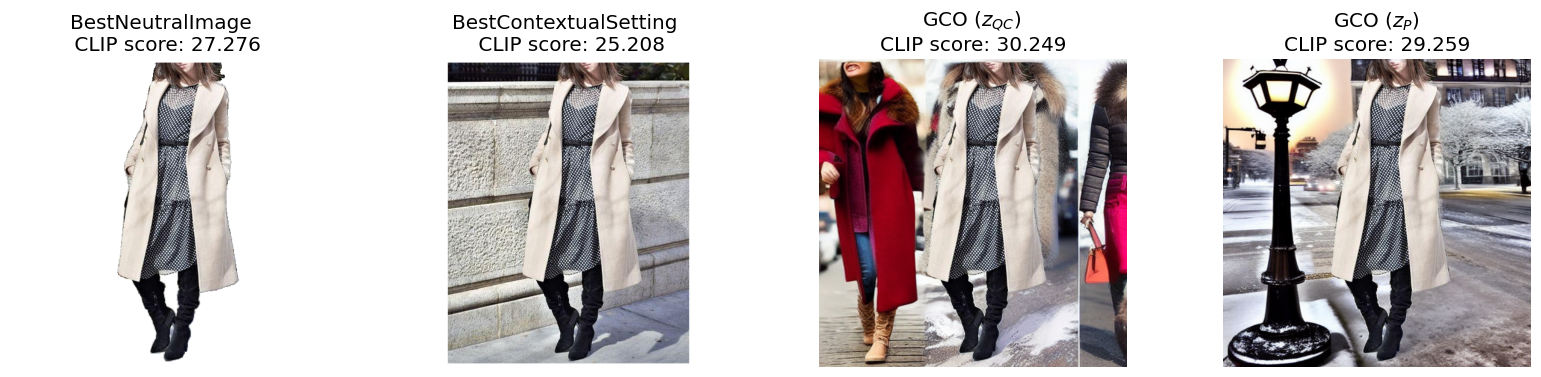}
\caption{Winter}\label{fig:winter_fail1}
\end{subfigure}

\begin{subfigure}[b]{\linewidth}
\includegraphics[width=\linewidth]{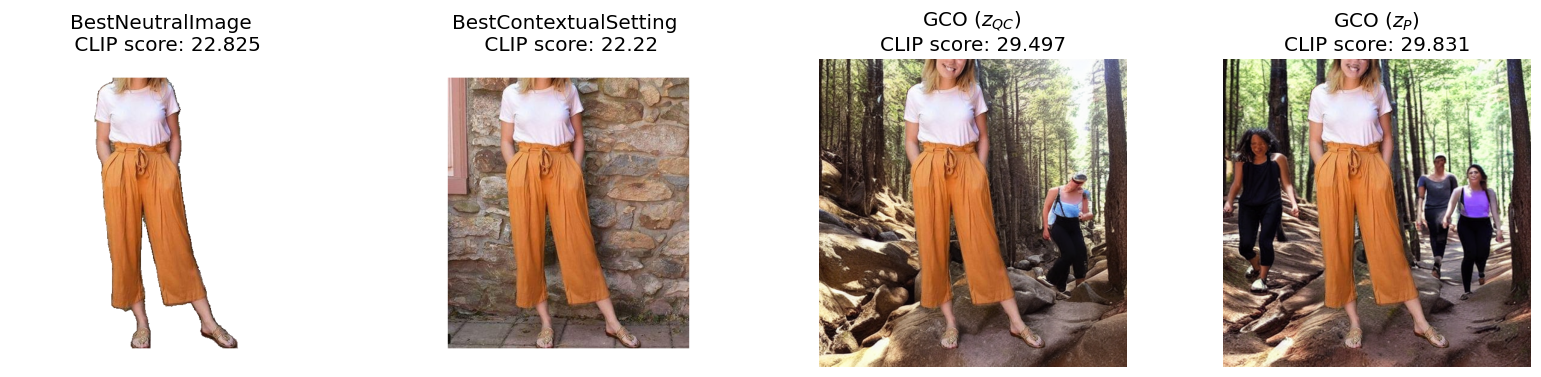}
\caption{Setting}\label{fig:setting_fail2}
\end{subfigure}

\begin{subfigure}[b]{\linewidth}
\includegraphics[width=\linewidth]{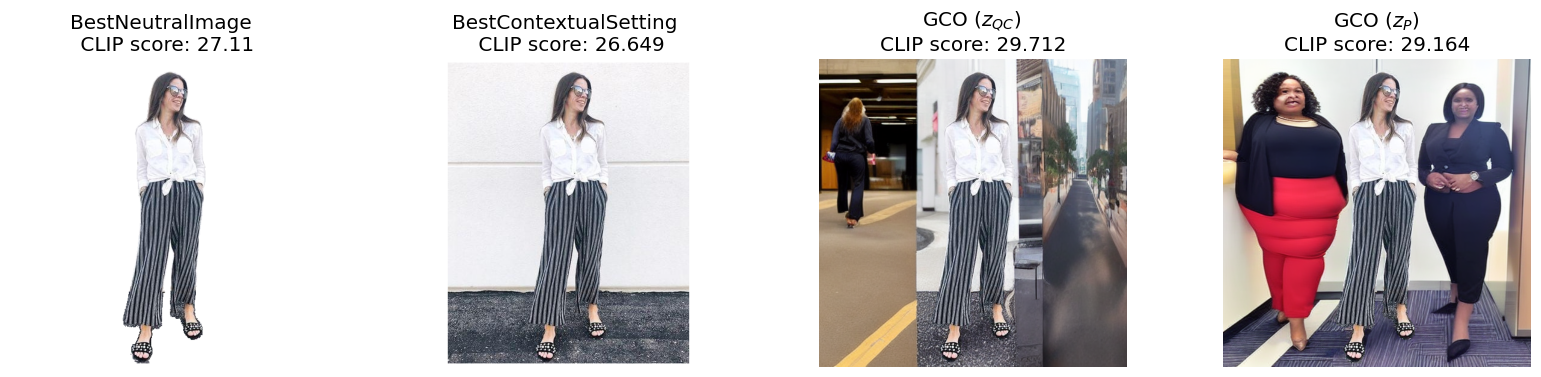}
\caption{Setting}\label{fig:setting_fail3}
\end{subfigure}

\begin{subfigure}[b]{\linewidth}
\includegraphics[width=\linewidth]{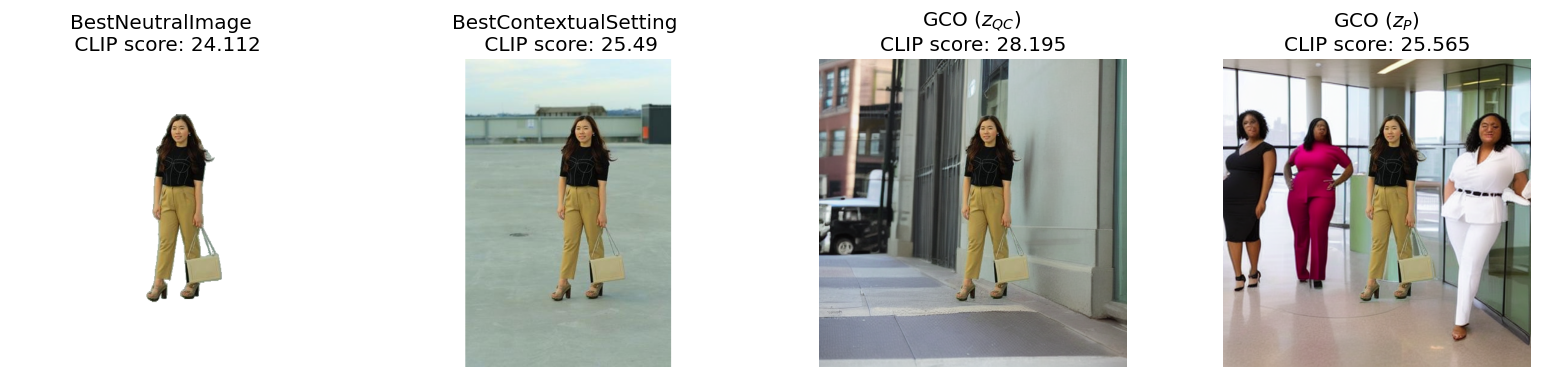}
\caption{Setting}\label{fig:setting_fail4}
\end{subfigure}

\caption{Examples of failed AdBooster model generations compared to predefined product images in different categories.}
\label{fig:seasons_fail1}
\end{figure}

Figure \ref{fig:seasons_fail1} depicts some occasional failures in \emph{GCO} images: poor quality of people generation, deduplicated images, problems with light casting and object/subject ratio. The corresponding $z_{QC}$ (Query, Context) and $z_{P}$ (Prompt) can be found below.

For figure \ref{fig:winter_fail1}:
\begin{itemize}
    \item \textbf{Query}: winter coats for women;
    \item \textbf{Context}: wear during the snowy winter season in the city;
    \item \textbf{Prompt}: Amidst a snowy cityscape, woman in a winter coat with a glowing streetlamp casting a warm and welcoming light over the scene.
\end{itemize}

For figure \ref{fig:setting_fail2}:
\begin{itemize}
    \item \textbf{Query}: comfortable hiking pants;
    \item \textbf{Context}: for a weekend hiking trip with friends;
    \item \textbf{Prompt}: A group of friends walk confidently through a picturesque forest trail, clad in comfortable and versatile hiking pants, enjoying the beauty of nature on a perfect weekend getaway.
\end{itemize}

For figure \ref{fig:setting_fail3}, \ref{fig:setting_fail4}:
\begin{itemize}
    \item \textbf{Query}: casual women’s pants;
    \item \textbf{Context}: wear to work at the office in the big city;
    \item \textbf{Prompt}: A group of professional women with diverse body shapes and skin tones are posing confidently in a modern, sleek office building lobby, wearing stylish and comfortable work pants that perfectly complement their professional attire.
\end{itemize}

% Figure \ref{fig:seasons_fail1} gives an example of both poor quality of people generation and of deduplicated image (\emph{GCO} $(z_{QC})$). We can also observe some minor problems with light casting and streetlight ratio with respect to depicted woman in \emph{GCO} $(z_{P})$ image.

\end{document}